# Deep Reinforcement Learning (DRL)-based Methods for Serverless Stream Processing Engines: A Vision, Architectural Elements, and Future Directions

Maria R. Read[1], Chinmaya Dehury[2], Satish Narayana Srirama[3] and Rajkumar Buyya[1]

[1]Cloud Computing and Distributed Systems (CLOUDS) Lab,
School of Computing and Information Systems,
The University of Melbourne, Australia

[2]Institute of Computer Science,
The University of Tartu, Tartu, Estonia

[3]School of Computer and Information Sciences,
The University of Hyderabad, Hyderabad, India

**Abstract:** Streaming applications are becoming widespread across an extensive range of business domains as an increasing number of sources continuously produce data that need to be processed and analysed in real time. Modern businesses are aggressively using streaming data to generate valuable knowledge that can be used to automate processes, help decision-making, optimize resource usage, and ultimately generate revenue for the organization. Despite their increased adoption and tangible benefits, support for the automated deployment and management of streaming applications is yet to emerge. Although a plethora of stream management systems have flooded the open source community in recent years, all of the existing frameworks demand a considerably challenging and lengthy effort from human operators to manually and continuously tune their configuration and deployment environment in order to reach and maintain the desired performance goals. To address these challenges, this article proposes a vision for creating Deep Reinforcement Learning (DRL)-based methods for transforming stream processing engines into self-managed serverless solutions. This will lead to an increase in productivity as engineers can focus on the actual development process, an increase in application performance potentially leading to reduced response times and more accurate and meaningful results, and a considerable decrease in operational costs for organizations.

## 1. Introduction

Traditionally, when developing and deploying applications in the cloud, users are required to address several concerns regarding the compute resources running behind the scenes. To mitigate the shortcomings associated with this, cloud service providers such as Amazon introduced function-as-a-service offerings via products like AWS Lambda. Although function-as-a-service is one of the first offerings to realise serverless computing in cloud environments, the serverless paradigm continues to evolve and encompasses a broad range of cloud services that aim to completely hide the complexity of infrastructure and operational management from application developers [1].

Serverless computing allows developers to focus on application implementation and innovation as they no longer have to deal with server administration, performance tuning, cluster management, and autoscaling, among others [2]. These responsibilities are delegated to the serverless framework, which receives the application code and automatically deploys it in a cloud environment. Relying on this model enables organizations to significantly reduce infrastructure costs, release revenue-driving applications more quickly into production, and have expert personnel focused on developing applications that create value for the organization.

Streaming applications are at the core of many modern businesses. For example, the manufacturing industry relies on streaming data to enhance production lines by monitoring performance and predicting faults before they occur [8]; real time network traffic analysis is used to identify and respond to security incidents [3]; and financial institutions are processing streams of data to detect changes in stock prices and automatically rebalance portfolios [4]. Health, social and environmental use cases are also abundant. For example, physiological data streams of patients undergoing anaesthetic can be analysed to raise early warnings [5], data generated during and after a disaster can be used to produce warnings through social media platforms [6], and live traffic data can be used to optimize traffic flow [7].

Streaming applications are represented as graph topologies with multiple inter-related components. Deployed topologies usually have a complex network of replicated components that are continuously processing vast amounts of data with stringent performance requirements such as millisecond end-to-end latencies [8][38]. They are long-running applications as their lifetime can span days, weeks, or even months. In fact, many are required to be deployed for an unlimited period of time, just like user-facing web applications. Stream Processing Engines

(SPEs) or systems [37] are large-scale distributed systems that are designed specifically with the characteristics of streaming applications in mind. Data tuples originating from a stream source flow through the system as they are generated, allowing real time data analysis without the need of storing data and processing it in batches [14]. There has been a considerable influx of SPEs into the open source community in recent years due to the rise in demand for real-time data analysis. Examples include Storm [9], Samza [10], Heron [11], Flink [12], and Spark [13]. Still, none of these systems has been designed with the purpose of simplifying their use from the application developer's perspective and much of the underlying complexity of the distributed system and the distributed infrastructure has been passed on to users to manage.

As a result, the use of existing SPEs is non-trivial and application deployment entails a considerable effort from software engineers who must define values for a myriad of performance-affecting configuration parameters and estimate the capacity of the compute infrastructure where the SPE will run [15]. More importantly, choosing appropriate configuration values to achieve a desired level of performance demands a thorough understanding of the internal workings of the SPE, the infrastructure in which the system is deployed, the dynamics and implementation details of the application's topology, and the rate at which data is generated by the source. This process also entails multiple manual trial and error runs repeated continuously throughout the lifetime of the application, as external factors such as load spikes affect its execution. Hence, current SPEs are prone to SLA violations, require constant expert technical personnel on call attending to incidents, and have a suboptimal use of resources, all of which leads to increased operational costs and decreased application performance.

We identify challenges in three levels of tuning, currently expected from users by existing SPEs:

*System Parameter Tuning:* Existing SPEs provide users with a plethora of configurable parameters, leaving it up to operators to manually explore different value combinations. Furthermore, they leverage container management systems to manage cluster resources. This brings another challenge for users, who must define the amount of CPU and RAM that are to be used by the application components, which is challenging as this requires extensive profiling of the application and workload forecasting. Values for these parameters have an effect in performance and cost: over-estimating resources results in higher infrastructure costs, while under-estimating results in reduced application throughput and increased end-to-end latencies.

*Application Parameter Tuning:* SPEs require users to specify the parallelism, or number of replicas, of each component in a topology. More replicas will enable components to process higher volumes of data, resulting in higher throughput. Underestimating this value leads to a topology riddled with backpressure, while overestimating it leads to low resource utilization. Furthermore, the parallelism of components must be adjusted throughout the lifetime of the application to compensate for failures, performance variation in compute and network resources, and load spikes.

*Infrastructure Management:* Clusters supporting the execution of SPEs can be deployed on any distributed environment, but in today's cloud-dominated era, this is almost always a public Infrastructure as a Service provider. Currently, no SPE provides cluster autoscaling capabilities to support changing application workloads. Instead, users must estimate the capacity of the cluster and i) overprovision resources to account for unforeseen load spikes or ii) manually adjust the cluster size as a reactive measure to a load spike. Overprovisioning results in reduced return on investment for an organization and under-provisioning may result in loss of revenue due to application downtime.

Overall, delegating the responsibility of these challenges to application and system engineers is inefficient, error prone, and time consuming. It is unrealistic to expect operators to have a comprehensive understanding of all the configurable system knobs and explore an immense search space to find a combination that achieves the SLOs for the current state of the system and underlying infrastrucure. Therefore, by leveraging multi-agent reinforcement learning, our proposal aims to offer SPEs the ability to automatically adapt in response to environmental changes in order to maintain the desired quality of service requirements. Figure 1 illustrates our vision for serverless SPEs (on the right) compared to traditional architectures (on the left). Along with the application's topology, users will be able to submit their desired SLOs (e.g., throughput), and delegate the responsibility to the self-tuning system to determine the system's and infrastructure's optimal configuration. Our proposed system will have the ability to react to external shocks by automatically scaling and reconfiguring itself as needed. It will automatically recover not only from failures but also from degradation of system performance.

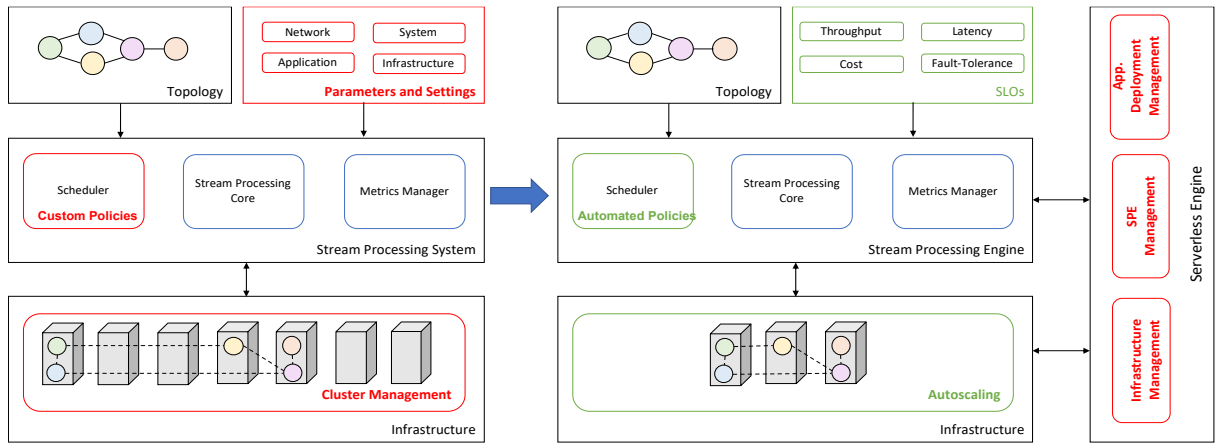

**Figure 1:** Architectural model from traditional to serverless stream processing engines.

## 2. Significance and Innovation of our Vision

Our vision article proposes to create serverless stream processing engines with autonomic capabilities. The proposal is innovative, original, and highlights significant problems that need to be overcome:

- **Timeliness:** Despite their increased adoption and tangible benefits, support for the automated deployment and management of streaming applications is yet to emerge. In today's x-as-a-service era, serverless computing is taking the cloud computing world by storm, organizations with stream processing needs are also seeking for ways to leverage the benefits of infrastructure-unaware and cloud-managed deployment. The demand for this type of solutions can be seen by the increased adoption of "serverless streaming services" offered by many cloud providers. Such approach to stream processing is not based on engines specifically designed with the needs of streaming applications in mind; instead, it is based on the integration of existing serverless models such as function-as-a-service (FaaS) (e.g., AWS Lambda) and backend storage-as-a-service (e.g., DynamoDB). The existing FaaS model constrains applications to be stateless, forcing topology components to manage state via remote storage systems, which greatly impacts performance. Another disadvantage of existing serverless stream processing deployments is the lack of support for complex topologies. Distributing data to different components based on a tailored function is non-trivial and requires additional development effort. Furthermore, using mainstream serverless components misses the opportunity to optimize application deployment by taking into consideration its very specific set of requirements and stringent performance needs. Therefore, although current "serverless streaming services" deliver in their promise of automated deployment, the scope of streaming applications that can benefit from them is greatly limited by the restrictive serverless model they are based on and the simplicity of their data processing graph. Our vision aims to transform purposely designed stream processing engines into serverless solutions, moving the difficulties of infrastructure management, deployment, and performance tuning away from operators. This will enable the essence of serverless computing to be met by eliminating infrastructure management responsibilities and will ensure that QoS requirements are automatically satisfied and maintained throughout the application's lifetime. In particular, by automating real-time data processing systems, these will become more resilient to equipment failure and human errors. The self-adapting system we propose will adapt and recover from failures affecting the performance of applications and, since the goal is to reduce human interaction, human errors will be less likely..

- **Originality and Innovation:** New advances in Deep Reinforcement Learning (DRL) have unlocked the potential of artificial intelligence to a range of domains with complex problems that need solving. The super-human performance obtained by many algorithms in AI benchmarking applications, such as Go and Atari games, suggest that the potential of these techniques must at least be explored in different areas. Serverless and stream computing are mainstream technologies underpinning countless modern businesses and applications, yet there is still vast room for improvement in them. They can greatly benefit from the ground-breaking and rapidly evolving DRL field. Exploring the implementation of a Multi-Agent DRL (MADRL) approach to combine them is novel compared to existing approaches automating systems from a single isolated perspective (e.g., scheduling, autoscaling, parallelism) based on traditional optimisation approaches such as greedy heuristics and evolutionary algorithms. Furthermore, enhancing SPEs with serverless capabilities as proposed in this vision aricle is also an innovative idea. It breaks away from the already defined and restrictive serverless model offered by cloud providers and acknowledges that it is insufficient to fulfil the requirements of streaming applications.

- **Advancement of the Discipline:** Exploring the intelligent automation of SPEs via DRL is a novel approach that has the potential of furthering not only SPEs, but the cloud and serverless industries as well. Just as SPEs, there are many distributed systems whose adoption is rising in cloud and edge/fog environments. They have similar requirements to streaming systems and as distributed applications, can greatly benefit from a better and smarter automation. The proposed vision will help in advancing many modern applications by demonstrating the applicability of DRL to the automation and resource management optimization problem in large-scale distributed systems.
- **Significant Business Opportunity:** Cisco estimates 500 billion devices to be connected to the Internet by 2030 [20]. This exponential growth gives place to a new digital era with unprecedented opportunities for organizations around the world. Devices such as surveillance cameras, cars, farm machinery, smart TVs, toys, and wearables, are examples of appliances that are part of this interconnected era and are continuously producing data. It is estimated that these devices will produce approximately 5 quintillion bytes of data everyday [21]. The analysis of these data is essential to businesses and people to deliver insights that enable them to make better and more informed decisions. Real time data analysis is therefore becoming critical for organizations to create, improve, and maintain their competitive edge. The challenge is to build the right digital infrastructure and enable the right set of applications to harness this data and draw insights. As prtoposed vision focuses this particular problem, it will create significant business opportunities for organizations by making stream processing engines more accessible, efficient, and easier to use. Not only will it do so in the stream processing field, but the ideas from this vision article transferable to the serverless paradigm in general, contributing towards the goals of the next-generation cloud computing where cloud resources are better utilized, infrastructure costs are reduced, less energy is consumed, and application management and deployment is fully automated.

## 3. Architectural Framework

We propose the use of Deep Reinforcement Learning (DRL) to optimise the values of different parameters across three different configuration levels (system, application, and infrastructure) and determine the set of actions that can be taken to solve performance problems that arise in the deployment of streaming applications. Despite recent advances in DRL, this approach is yet to be explored in the context of self-tuning systems and SPEs. The steps for realization of proposed vision consists of the following:
- Architectural framework for serverless SPEs with the ability to maintain desired SLOs;
- Multi-agent DRL approach that enables the automated management of SPEs and cluster resources;
- Policies that reduce the exploration cost for RL agents in SPEs deployed in large-scale distributed platforms;
- Prototype system and demonstrator streaming applications deployed in a Cloud computing environment.

In a RL approach, feedback from the environment determines which action an agent will take next, with the agent making value judgements in order to select good actions over bad ones. In the context of this proposed vision, the RL agents will work towards a user-defined SLO goal, such as maintaining a desired throughput rate. The environment corresponds to the state of the deployed SPE, application, and compute infrastructure. The actions entail changes to the environment that affect the performance of the engine. These include scaling the cluster resources in or out, changing the parallelism of one or more components in the application, reconfiguring the placement of the components, and changing the values of the SPE's parameters, among others. A reward function measures the effectiveness of an action, or set of actions, towards the desired SLO.

There are multiple challenges when implementing the aforementioned DRL-based solution. First, the optimization space is broad and complex. To illustrate this, consider the case of tuning the parallelism of a topology with *5* components. Assume a cluster capable of accommodating a total of *50* component instances; that is, it has *45* free slots to accommodate replicas. Exploring the search space would involve exploring *49!/45!(4!) = 211876* different parallelism combinations. This over simplistic scenario is not considering the placement of components within the cluster machines, their communication overhead, or the heterogeneity of machines and topology components.

Second, the action space is also extensive. In the above example, actions to tune the parallelism of an application could be defined as increasing/decreasing the number of replicas of a component by one or leaving it unchanged. This would lead to $3^5 = 243$ discrete actions for a small topology with 5 components. The system should also be capable of considering actions related to the autoscaling of the cluster, the placement of components, the allocation of resources within a machine, and the values of system parameters. This leads to an action space of significant cardinality, which results in the framework requiring a considerable amount of learning time and data.

Third, the state of the system is broad and complex; it is represented by the machines in the cluster, the placement of operators within these machines, the resource utilization of operators, the values of the SPE's parameters, the internal state of the SPE such as the size of the input and output buffers, the data emission rate, and the value of QoS parameters such as throughput and latency. Modelling and capturing this state with minimum performance interference and determining what is meaningful for the agent to observe requires careful design and evaluation.

Fourth, balancing the exploration vs. exploitation trade-off is particularly challenging in distributed systems with real-time processing needs. In this article, agents have a restricted ability to freely explore their environment as this can significantly hurt performance (e.g., scaling down resources), can be time consuming (e.g., provisioning resources and redeploying the topology), and can result in additional costs (e.g., scaling out the cluster). Finally, defining functions capable of quantifying the efficiency of an action towards a desired goal is not straightforward. For example, although an action is appropriate to eliminate a performance bottleneck in a topology's execution, the performance benefits of such action may only be seen after a certain amount of time has passed and the system has reached a steady state. In fact, performance may first decrease as the topology is redeployed for example. Capturing this behaviour in the evaluation mechanisms guiding the reward process is a challenging task. With these challenges in mind, we propose a Multiple Agent Deep Reinforcement Learning (MADRL) system as depicted in Figure 2 to realize serverless stream computing, with the architectural components discussed in the following section.

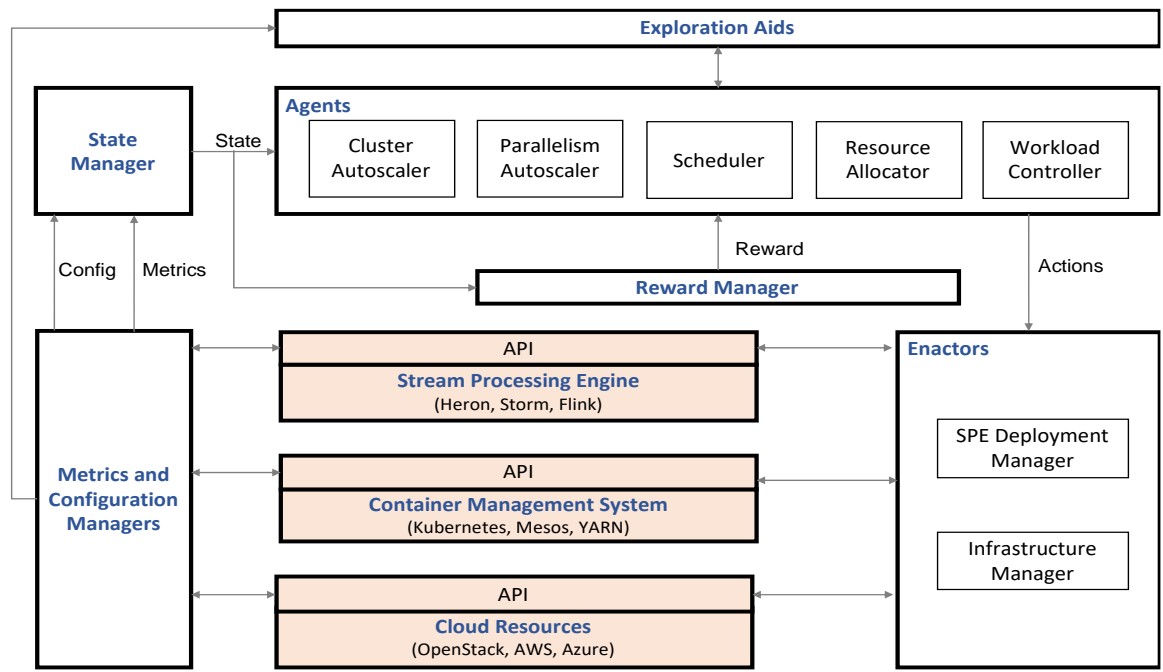

**Figure 2:** DRL-based Serverless Framework for Stream Processing Engines.

## 4. Research Issues and Envisioned Approaches/Future Directions
### 4.1 DRL-based Resource Management Agents
RL enables agents to learn by trial and error based on rewards obtained from their environment. In DRL, this knowledge is built using multi-layer neural networks, which progressively extract higher level features from raw input observed by the agents. This enables DRL to be applied to high-dimensional problems and hence, makes it a promising approach to solving complex real-world problems. We propose the use of DRL to compute policies that drive the decision-making process of agents managing the deployment of streaming applications and tuning their performance. There have been considerable advances in DRL since 2015, when the concept of deep Q-networks (DQN) [23] was used to create an agent capable of outperforming professional players in Atari games. Following this, DeepMind created AlphaGo [26] and AlphaGo Zero [25], two agents combining DRL with Monte Carlo tree search capable of defeating professional players on the game of Go. More recently, DeepMind released the A3C algorithm [30], which has demonstrated better performance than DQN in different domains, is faster, simpler, and more robust.

Although both DQN and A3C can be used to train powerful agents, their performance may suffer when faced with complex tasks and when there are long delays between actions and reward signals. This is the case when reconfiguring an SPE; an improvement on the throughput of the application may only be seen after a while when the system has stabilized. Also, the model-free approach used by these algorithms, even though it leads to better performance, entails agents learning from scratch by trial and error. This may not be feasible considering the cost of reconfiguring the SPE in monetary and performance terms. Hence, investigating emerging approaches combining model-free and model-based techniques is necessary. This will allow agents to use a learned model to generate a simulated experience to help train a model-free policy. Examples of recent algorithms in this area are Imagination-Augmented Agents (AI2) [27] and Temporal Difference Model (TDM) [28].

Another important factor to consider in the design of the agents is that the goal of many DRL approaches is to be general enough so that they can be applied across a wide range of domains. As this approach can be unsuitable and as a result, refining algorithms so that they are tailored for the specific scenario of SPE auto-management will be necessary. Overall, applying suitable DRL concepts and adapting them when necessary are key challenges for this critical component of the proposal.

Also, a single-agent DRL solution entails training a single policy that issues all the actions. Considering the complexity and nature of the given problem, a multi-agent RL solution offers a more natural decomposition of the problem, the potential for more scalable learning, and the flexibility to extend the framework with more agents later on. Instead of training a single, super-agent that controls all different aspects of performance such as autoscaling, parallelism, and parameter tuning, we propose using Multiple Agent Deep Reinforcement Learning (MADRL) to have multiple specialized agents focusing on different aspects of the system but working towards a common goal. Such decomposition of the action and state space also reduces the dimensionality of the problem, addressing two of the main challenges previously identified, and the size of the search and action spaces.

MADRL research supports the idea that collaboration and competition are the basics of collective intelligence and has strong connections with single-agent DRL, multi-agent systems, game theory, evolutionary computation and optimization theory. DeepMind made another breakthrough in their contribution to AI by publishing their experiences building collaboration and competition in DRL agents playing Capture the Flag game [31]. Recently, Jaques et al. [29] have extended work in the area by proposing an algorithm that allows collaborative and competitive dynamics between DRL agents to emerge in ways that even surpass human counterparts. They propose an architecture in which agents are rewarded for having a casual influence on other agents' actions, which allows them to learn to cooperate. Their architecture proposes a network that enables an agent to predict the actions of every other agent, allowing the agent to simulate and predict how its actions will affect other agents. This eliminates the need for centralized training or the sharing of reward functions, making it an appealing work to explore as part of the implementation of our multi agent system. Given that computational efficiency is essential in our system, we propose to improve the cooperation efficiency between multiple agents by learning a causal graph between multiple agents instead of using a fully connected graph.

In our MADRL approach, agents work towards a common goal. They learn the dynamics of the environment by trial and error while cooperating with other agents. In particular, we propose a system with the following agents:

- *Cluster Autoscaler:* This agent is responsible for managing the capacity of the cluster where the SPE is deployed. This entails removing virtual machines if the utilization of resources is low for example. If the cluster is overloaded and resources are needed to eliminate a bottleneck in the application, then the agent should launch new VMs to better cope with the current load. Overall, the goal of this agent is the efficient management of the cluster resources with the aim of increasing their utilization and reducing cost while maintaining the desired SLOs.
- *Parallelism Autoscaler:* Tuning the parallelism of a topology is a challenging task. The relationship between components means that altering the parallelism of a component will almost always require adjusting the parallelism of upstream and downstream components for it to have a meaningful impact in the overall performance of the application. This agent will automate this process by learning the best set of actions to take when a lack of parallelism is affecting the performance of an application.
- *Scheduler:* The agent selects an available resource to deploy a particular instance of a component while considering user-defined objectives. For example, for throughput maximization, components that communicate with each other should be co-located in the same hosts. This however is unachievable in a large-scale system and as a result, components should be placed so that the communication overhead is minimised. Different placement policies can be explored in this module as issues like resource utilization, load balancing, data locality, and resource consumption of existing containers can be considered.
- *Resource Allocator:* Allocates a specific amount of resources to components of a topology. These resources usually refer to the amount of RAM and CPU an instance of a component will use. The agent may decide to increase or decrease resources if this means achieving the desired SLOs. For instance, backpressure in a topology may be due to a component not having enough resources to process tuples fast enough. Increasing these may result in the topology being back in a healthy state.
- *System Parameter Tuner:* This agent is responsible for tuning the configurable parameters of the SPE, such as queue capacities, socket buffers, network packet sizes, and read/write batch sizes among others.

**4.2 Exploration Aids**

Applying DRL algorithms to real-world problems can be challenging; even when the task of the agent is well-defined and there is an explicit reward function. Often, it is not possible to let an agent interact freely with its environment due to external constraints such as cost and time. In order to deal with these limitations, it is important to consider different methods that help agents explore their worlds more efficiently.

An approach that needs to be investigated to solve this problem is to enhance the RL solution with efficient initialization policies that reduce the learning time for online decisions. Exploration aids will then use data collected from the execution of applications with typical configurations and use these sample data to train models that allow the agent to predict the performance of applications under similar configurations. For example, runtime metrics can be collected by running a variety of applications on a given SPE and executing random or deliberate actions to the system. This data can then be used to train and initialize the actor and critic neural networks used by agents. Based on this initial idea, we will explore counterfactual reasoning methods to augment the observational data by sampling from a causal generative model [32]. Another option that will be considered here is to use a simulator that mimics the real environment and train the agents in it. We will develop deep transfer learning methods that enable the agents to transfer from the simulation environment to the real-world environment. One promising idea is to incorporate an encoder that extracts environmental information, which can be used to develop environment-adaptive policies. The other way is to learn invariant representations that have similar distributions on the simulated and real-world data. In an RL setting where gathering experience in the real-world is expensive and constrained by performance requirements, this approach can be used to apply a policy learnt in a simulation environment in a production deployment.

An additional methodology to be explored in this area is the modelling of the dependencies between different configuration parameters and the use of this knowledge by the agents. It has been observed in different contexts [19] [22][23] that the relationship between parameters plays an important role when looking for an optimal configuration. This information may aid in reducing the number of reconfigurations done by agents before achieving a steady state, reducing the exploration time and cost of the proposed solutions. Evaluating the usefulness of these techniques and determining their effect on the performance of the agents will be an important outcome of our vision.

### 4.3 Policies for Management of Rewards, States, and Enactment

Given an observation of the system's state, the reward manager applies different evaluation mechanisms that accurately capture and quantify the efficiency of the corrective actions taken by the agents. In RL settings, reward functions define the goals that need to be achieved by the agents. This can be a rather challenging task due to the complexity of the environment as defined in this article. For example, a scheduling agent needs to be able to produce conflict-free environments that minimize the communication overhead between components. The reward values need to reflect these two objectives as the agent ultimately has a single goal, maximizing the expected reward.

The *State Manager* accesses relevant metrics and configuration settings and periodically estimates the system's state. The system's state may be expressed in terms of higher-level metrics such as tuple arrival rate, current cost of the leased infrastructure, tuple throughput, application latency, existence of data skew, etc. Furthermore, the state can include configuration settings relevant to the agents' goals such as placement configuration, resource allocation, and cluster capacity. The state can also be split into multiple partitions, each of which may be relevant to a particular agent. This may help in reducing the observation space for individual agents, allowing the RL algorithms to perform better and faster. Overall, the role of the *State Manager* within the architectural framework is to enable the seamless integration of different state management techniques and to decouple the agents from the raw metric collection process. This eliminates the need to restructure or re-implement the *Metrics Manager* if different observation patterns are being trialled when developing the agents. The realization of this module requires modelling the state of the system. This is a key task that will not only have a direct impact on the behaviour of the DRL agents but will also determine which algorithms are suitable to solve the problem at hand, this includes defining the dimensionsionality of the state as well as its continuous or discrete representation. Furthermore, this module requires the development of techniques to efficiently collect, manage, and aggregate disparate data from heterogeneous sources.

In our proposed framework, enactors are responsible for carrying out actions as determined by different agents. The actions can alter the underlying infrastructure, the allocation of resources, and the placement of components through the *Infrastructure Manager* and the configuration of the streaming application through the *SPE Deployment Manager*. To achieve this, we will develop techniques for seamless and automatic execution of actions in distributed environments with minimal effect on performance. To adjust the parallelism of a topology, we will develop algorithms that capture the state of the application at runtime and the load of the components' instances to be redistributed without losing the current state of the application and redeploying the entire topology.

### 4.4 Automated Techniques for Metrics and Configuration Management

The *Metrics Manager* collects performance, health, load, and other metrics from the SPE, the container management system, and the cloud provider. These time series data are stored in a repository to be accessed by the Sate Manager and Exploration Aids modules. Examples of measurements recorded from the SPE include input and output queue sizes, time spent in garbage collection cycles, whether back pressure has been detected, number of acknowledgements, etc. The container management system can be queried for the cluster capacity, state of tasks, and resources consumed. From the cloud provider, data such as the lease period of virtual machines will be obtained.

The *Configuration Manager* keeps track of the current system configuration and shares this information with the *State Manager*. An example of a configuration that is relevant to the system is the current mapping of components to machines. This information should be included as part of the state reported to an agent trying to optimize how a topology is scheduled in the cluster to minimize communication over the network. The agent would then be able to associate a given mapping with a certain communication overhead to take appropriate actions towards its goal.

**5. A Case Study - Use Case Scenario: Multi-source Stream Video Analytics System**

**5.1 Use case overview**

**Video Feed Sources**: To demonstrate the envisioned architectural framework, a multi-source streaming video analytics application is considered. This application serves as a comprehensive video analytics platform capable of ingesting video data from multiple sources. For instance, several cameras in a "Smart Traffic Monitoring" system are possibly placed at traffic intersections or roads, diligently monitoring the flow of traffic, potential violations, and other vehicular activities. CCTVs in "Surveillance" systems could be strategically placed in public spaces, commercial premises, or even private properties. Video may also originate from "Aerial Imagery Footage," where cameras are fixed on aerial vehicles like drones, capturing extensive footage from an elevated or bird's-eye view. The core functionality revolves around preprocessing, conditional service distribution, video analysis, and storage.

**Pre-processing and Conditional Distribution**: The incoming stream of video needs to be preprocessed. The preprocessing stage includes ensuring consistent video formats and compatibility with downstream processes, shrinking video sizes by dropping unnecessary frames or reducing frame rates, and ensuring videos adhere to specified width and height specifications. The preprocessed videos are further forwarded to a *conditional service distribution* to route the preprocessed videos to different services based on predefined conditions. The examples of predefined conditions can be (a) based on time, where videos are processed at regular time intervals, (b) based on the size of the video, where larger videos might be treated differently from smaller videos, or (c) based on the source of the video.

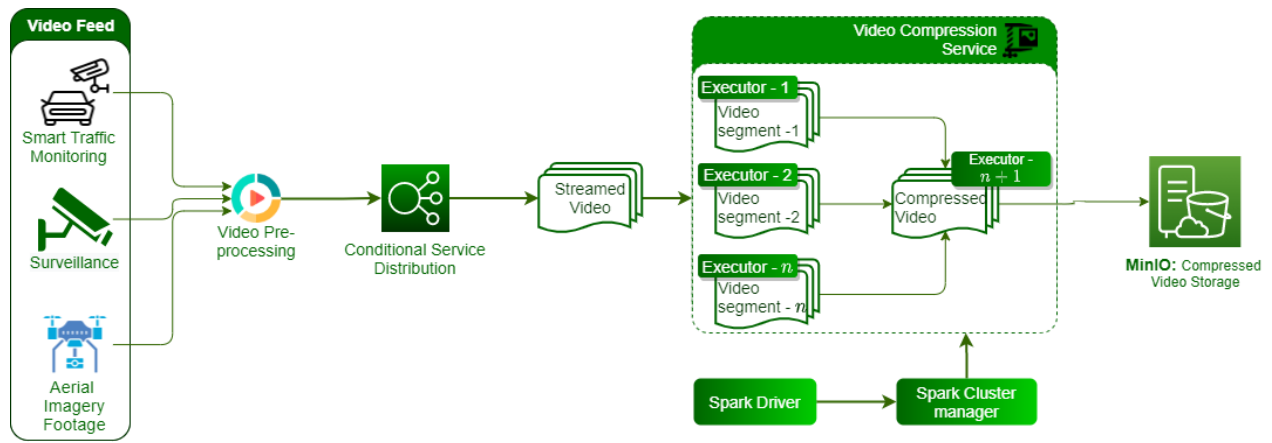

**Figure 3**: Use case overview: Streaming Video Processing

**Video Processing**: The same video can be sent to multiple independent services at the same time. In the current experiment, one service is implemented. This streamed video then undergoes a significant transformation in the "Video Grayscale Service". Here, the original video stream is converted into grayscale. The rationale behind this is multifold – from conserving storage space and enhancing processing speeds to focusing primarily on detecting movement patterns and shapes instead of varied color palettes.

The implemented service uses Apache Spark[1] and its streaming-related components. Apache Storm[2] and Apache Spark are two popular open-source stream processing engines that provide near-real-time latency. Both engines are highly scalable and process large volumes of data in a timely manner. In this experiment, Apache Spark is used over Apache Storm. This is primarily due to its ability to process data using in-memory computation, which makes it faster than Apache Storm [33, 34, 35]. In-memory computation primarily refers to storing and processing data directly in the RAM of the distributed nodes across the cluster, unlike traditional disk-based processing, where data needs to be read/write from/to a disk. Additionally, Apache Spark is easier to program and fault-tolerant than Apache Storm, due to its resilient distributed datasets (RDDs) and dataset API. The community involved in Apache Spark is more active than Apache Storm's community. This can be observed from the frequency of contributions to their respective Github repositories.

---
[1] https://spark.apache.org/
[2] https://storm.apache.org/

At this stage, the streaming video frames are processed by multiple Spark executors parallelly for faster processing. These units (executors) work in parallel, each handling distinct segments of the video feed. Such a division allows the system to achieve heightened processing speeds and efficiency. Each executor uses the FFmpeg[3] Python library for video-related operations. Upon grayscale conversion, the transformed video is labeled as 'Processed Video' and is handled by a dedicated executor.

**Distributed Computing with Spark and Storage**: Figure 3 shows two crucial components related to Spark: the Spark Driver and the Spark Cluster Manager. The "Spark Driver" is the central orchestrating process, coordinating various tasks. It is responsible for breaking down the application into smaller chunks that can be distributed across VMs in the cluster, ensuring that tasks are dispatched to available executors in an optimized manner, aggregating the results, and producing the final output. On the other hand, Spark Cluster Manager is responsible for managing the underlying VMs on which the application runs, allocating resources (e.g. CPU and memory) based on the requirements of the tasks or their configuration, monitoring the health of the VMs and re-allocating tasks in case of VM failures. Once the video segments have been processed through the Spark framework—divided, processed in parallel, and aggregated—the results are stored in the MinIO-based[4] storage system. MinIO provides an object storage solution that is scalable and compatible with the Amazon S3[5] API, making it a suitable choice for storing large volumes of video data.

### 5.2. Use case implementation

Figure 4 shows the implementation architecture of the proposed use case on streaming video grayscale and the underlying infrastructure.

### 5.2.1. Infrastructure Configuration and Toolstack

**Table 1:** Deployment infrastructure using Kubernetes

| VMs | Role | OS | Flavor | Storage | Security Group |
| --- | --- | --- | --- | --- | --- |
| Master | Controller, worker | Ubuntu 20.04 | 4 vCPU, 8 GB RAM | 30 GB | allow-all |
| Worker1 | Worker | Ubuntu 20.04 | 4 vCPU, 8 GB RAM | 30 GB | allow-all |
| Worker2 | Worker | Ubuntu 20.04 | 4 vCPU, 8 GB RAM | 30 GB | allow-all |
| Worker3 | Worker | Ubuntu 20.04 | 4 vCPU, 8 GB RAM | 30 GB | allow-all |
| Client | User | Ubuntu 20.04 | 2 vCPU, 4 GB RAM | 25 GB | allow-all |

The experiment relies on the computing and storage resources provided by the High Performance Computing Center at the University of Tartu[6] [C1]. The infrastructure consists of four VMs (as shown in Table 1): one acts as the client and the remaining three form the computing cluster. Ubuntu 20.04 Operating System (OS) is installed in the client VM, and it is equipped with 2 vCPUs, 4GB of RAM and 25GB of storage. The client VM serves as a management and user interface for the entire system. The Kubernetes cluster consists of three VMs: *master*, *worker1*, *worker2*, as shown in Table 1. Each VM is configured with Ubuntu 20.04 OS, 4 vCPU, 8GB of RAM, and 30GB of storage.

---

[3] https://pypi.org/project/ffmpeg-python/

[4] https://min.io/

[5] https://aws.amazon.com/s3/

[6] https://hpc.ut.ee/

As shown in Figure 4, the use case is deployed atop a Kubernetes cluster and a number of tools are used for the overall management. Table 2 shows the list of tools used in this implementation, including their version and purpose in this implementation.

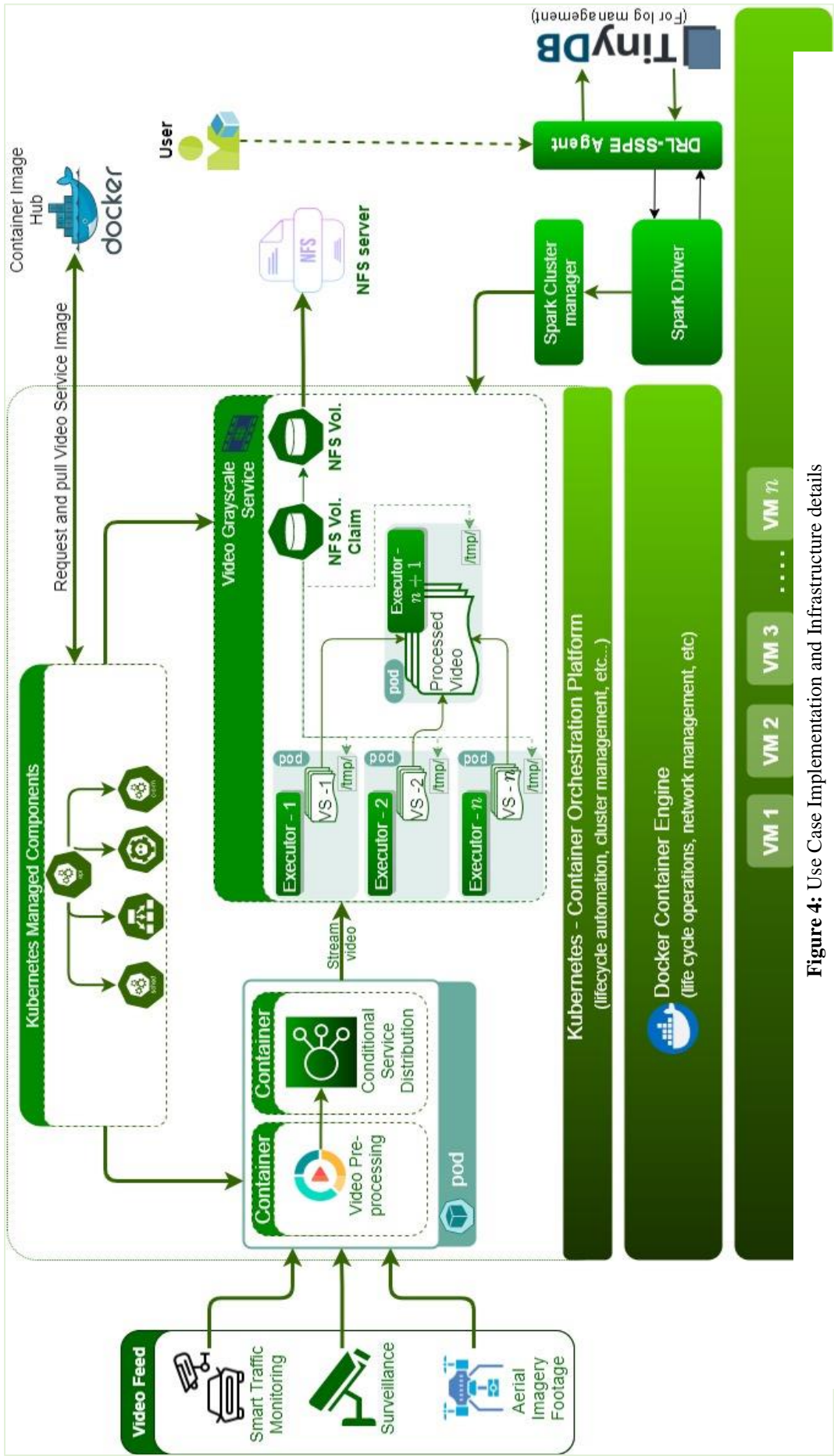

**Figure 4:** Use Case Implementation and Infrastructure details

Table 1: Toolstack used in implementation

| Tool Name | Version | Purpose |
|---|---|---|
| Kubernetes (k3s) | 1.27 | An open-source container orchestration system for automating application deployment, scaling, and management. Kubernetes is used to manage and deploy Docker containers that are created for spark executors. |
| Docker | 24.0.5 | A platform used to develop, ship, and run applications inside containers. This is used to package up the video grayscale application and its dependencies together. The developed container image for video grayscale is stored in Docker Hub [7]. This ensures that the application runs consistently across different environments. |
| Apache Spark | 3.4.1 | An open-source, distributed computing system that can process the large amounts of incoming video from multiple sources. It also provides an interface for programming entire clusters with implicit data parallelism and fault tolerance. |
| Tensorflow | 2.14.0 | A popular open-source ML library that comes with Keras API. Keras' simplicity, modularity, and extensibility features are used to develop our DRL model. |
| OpenCV | 4.8.0 | An open-source computer vision and ML software library for pre-processing and processing the video. |
| NFS | v4 | NFS (Network File System), a distributed file system protocol that allows one executor to access video data over a network as if the network devices were attached to its local file system. |
| TinyDB | 4.8.0 | A lightweight document-oriented database optimized for use in small-scale environments. This is used for logging the DRL agent's actions and execution times of video processing application. |

k3s[8] version 1.27 is used for simplifying the installation and configuration process of Kubernetes. The latest Apache Spark v3.4.1 is installed atop the Kubernetes cluster. Deploying Spark jobs on Kubernetes pods leverages the platform's capabilities for pod management, scaling, and resource allocation. This integration provides benefits such as dynamic resource allocation, enhanced Spark tasks' isolation, and seamless deployment in cloud-native environments. *Listing 1* shows the *spark-submit* command to deploy video grayscale service on Kubernetes cluster.

```
./bin/spark-submit  --verbose \
        --master k8s://http://172.17.67.89:8001    \
        --deploy-mode client    \
        --name spark-video-compress    \
        --conf spark.kubernetes.namespace=default \
        --conf spark.executor.instances=5    \
        --conf spark.executor.cores=2 \
        --conf spark.executor.memory=1000m \
        --conf spark.dynamicAllocation.enabled=true \
        --conf spark.kubernetes.container.image=apache/spark-py    \
        --conf spark.kubernetes.authenticate.driver.serviceAccountName=spark \
        --conf                      spark.kubernetes.executor.volumes.persistentVolumeClaim.nfs-pvc-volume.options.claimName=my-nfs-pvc \
        --conf                      spark.kubernetes.executor.volumes.persistentVolumeClaim.nfs-pvc-volume.mount.path=/tmp \
        --conf spark.driver.host=172.17.67.89 \
        ./myproj/drl_sspe/video_grayscaller.py
```

**Listing 1:** Code snippet for deploying the use case on Apache Spark using "spark-submit" command

---

[7] https://hub.docker.com/

[8] https://k3s.io/

**5.2.2. Implementation Description**

This section discusse\s the purpose of each component in demonstrating the video grayscale application, as shown in Figure 4. The demonstration leverages the strengths of Kubernetes and Spark within a private cloud framework. At the start, we have various video sources, such as smart traffic monitoring systems, general surveillance cameras, and aerial imagery footage from devices like drones. All these feeds provide real-time video data that is forwarded to the Kubernetes-managed environment for processing. Inside this environment, the raw video data first goes through the Video Pre-processing Service followed by the Conditional Service Distribution, which determines the route for these videos, as discussed before. Spark is tightly integrated into this architecture for distributed computing. Each executor is hosted inside a pod with one container. The executor accesses the "/tmp/" directory inside the container's file system to access the video chunk. NFS server is installed on one of the servers for the storage of video data. A persistent volume is created in Kubernetes environment with "nfs" storage class. Accordingly, a persistent volume claim is created for this application, which will mount the NFS directory to the "/tmp/" directory of each executor's pod. The video grayscale application is containerized and the build image is pushed to Docker Hub. The container image can be accessed with the name "*reachchinu/spark-py:opencv*". The container inside each pod is created based on this container image.

As discussed before, Spark's driver interacts with Spark's cluster manager for the overall management of application execution. However, the developed DRL-based algorithm "DRL-SSPE" interacts with Spark's Driver (through *spark-submit* command) for the resource configuration of the application. The configuration set provided by DRL-SSPE includes:
- *SPARK_EXECUTOR_CORES*: This refers to the number of CPU cores to be used by each executor. For instance, if SPARK_EXECUTOR_CORES=2, then each executor will use 2 CPU cores. This directly affects the number of tasks an executor can run concurrently.
- *SPARK_EXECUTOR_INSTANCES*: This defines the total number of executors that will run across the cluster for a particular Spark job. For example, setting SPARK_EXECUTOR_INSTANCES=10 means that 10 executor processes will be launched across the cluster nodes.
- *SPARK_EXECUTOR_MEMORY*: This specifies the amount of memory to be used by each executor. This memory allocation includes both the memory used by data structures inside the application and the overhead memory (memory used by Spark's internal operations). For instance, if SPARK_EXECUTOR_MEMORY=4g, each executor will get 4 gigabytes of memory.

DRL-SSPE also communicates with TinyDB for logging the cluster's performance history and events. This includes the configuration provided by DRL-SSPE, the number of video data partitions, the size of the batch (video chunk), the number of frames present in a batch, timestamp, execution time for each running instance, etc. TinyDB stores the data in JSON format. In the current implementation, the database can be accessed from the "/logs/" directory present in the Spark's working directory.

The DRL-SSPE agent deploys the Spark jobs from the client VM. Using the "*spark-submit*" command, the jobs are deployed in *client* mode, with the flag "*--deploy-mode client*", as shown in Listing 1. This indicates that the Spark *"driver"* is deployed on the client machines, while other tasks are deployed on the Kubernetes cluster, as shown in Figure 4. Figure 5 shows the detailed execution flow of DRL-SSPE during its training phase.

The process starts with the invocation of DRL-SSPE. This initiates the training process of the agent. The agent explores a different set of configurations (*SPARK_EXECUTOR_CORES, SPARK_EXECUTOR_INSTANCES, and SPARK_EXECUTOR_MEMORY*) based on its reinforcement learning model to ensure optimal performance. For each combination of parameters, the agent invokes a script (*deploy_application.sh*) which will eventually deploy the video grayscale application through *spark-submit* command as given in Listing 1. This also observes the execution time and other performance metrics. The agent further uses the metrics to update and optimize its policy.

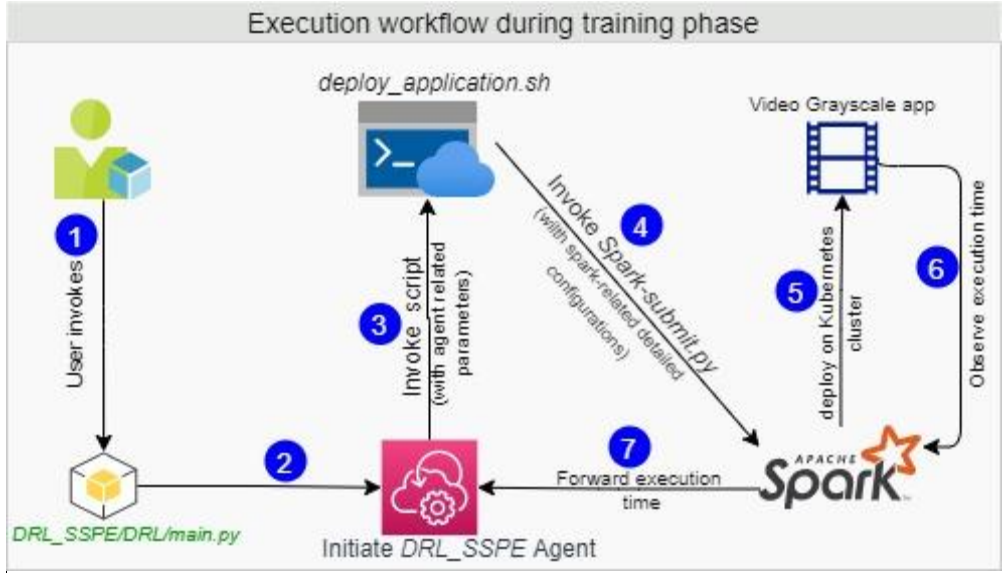

**Figure 5:** Execution workflow during training phase

**5.3. DRL-SSPE implementation for video grayscale**

This section discusses the design and development of DRL-based algorithm for optimal allocation of cluster resources for the Spark application. Creating a DRL algorithm to optimize the configuration of an Apache Spark application deployed on a Kubernetes cluster involves (a) defining the Environment (the state, action, and reward for the DRL agent), (b) the agent (implementation of the DRL agent using Deep Q-network) and (c) a training phase (training the DRL agent to find the optimal configuration). The reward value is inversely proportional to the processing time of the video frames. The processing time is measured in milliseconds. The goal of the DRL agent is to maximize the reward by minimizing the video processing time by tuning the configuration of Spark and the allocation of resources.

**DRL-SSPE implementation environment**:
The environment is the Kubernetes cluster where the Apache Spark job is being deployed and executed. Every time the DRL agent takes an action by changing configuration values, a new Spark job is submitted to the Kubernetes cluster, and the execution time is observed. Table 3 outlines the parameters that the DRL agent is responsible to tune based on the input video stream and other spark-related parameters.

**Table 2 : Apache Spark parameters**

| Resource Configuration | Min | Max |
|---|---|---|
| SPARK_EXECUTOR_CORES (number of CPU cores per executor) | 1 | 3 |
| SPARK_EXECUTOR_MEMORY (amount of memory per executor) | 500 | 1000 |
| SPARK_EXECUTOR_INSTANCES (Total number of executors) | 5 | 8 |
| Spark container image | reachchinu/spark-py:opencv[9] | |
| persistent volume type | nfs | |
| Max waiting time (maximum time to wait for a input video to finish processing) | 10 mins | |

Table 4 outlines other DRL and application-related parameters. The *Gamma* value, which reflects the discount factor for future rewards, is set at 0.95, suggesting a moderate emphasis on future rewards versus immediate ones. With an initial value of *Epsilon* = 1.0, the DRL agent will prioritize exploring the action space. For the exploration rate, we set the minimum value of Epsilon (i.e. *Epsilon_min)* to 0.01, ensuring that the agent retains a minimal level of exploration. The value of *Epsilon_decay = 0.995* will gradually reduce the epsilon value from its initial state to encourage a shift from exploration to exploitation of the learned policy. The Learning rate is set to 0.001, indicating a cautious approach to updating the agent's knowledge. The agent is configured to go through a maximum of 500 episodes, which are individual sequences of interaction with the environment. Related to the video partitioning for spark environment, the number of partitions ranges between 5 and 8. For each partition, we collect a minimum of 10 frames and a maximum of 60 frames before forwarding the frames to Spark environment for compression.

---

[9] https://hub.docker.com/repository/docker/reachchinu/spark-py/general

**Table 3 DRL and application-specific parameters**

| Parameter | Value | |
|---|---|---|
| Gamma | 0.95 | |
| epsilon | 1.0 | |
| Epsilon_min | 0.01 | |
| Epsilon_decay | 0.995 | |
| Learning rate | 0.001 | |
| episodes | 500 (max) | |
| Number of partitions | 5 (min) | 8 (max) |
| # of frames per partition | 10 (min) | 60 (max) |

### 5.4 Results Discussion

In the above sections, we discussed the implementation of a streaming video compression application using Apache Spark and presented detailed architecture. For a stream of video, the designed DRL algorithm determines the amount of resources to allocate to a given application and accordingly configures Spark to process the incoming video stream. In this section, we discuss the observed implementation results in terms of video processing time, the performance of the DRL agent, and the computation overhead incurred by DRL and Apache Spark.

#### 5.4.1. Video processing time

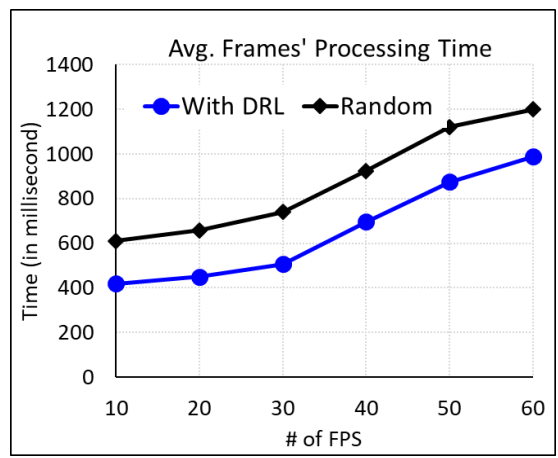

**Figure 6: Avg. Frames' Processing Time**

Figure 6 provides an empirical comparison between two resource configuration strategies for the Spark-based video processing application, as mentioned in previous sections. We observed the time taken to process varying numbers of frame rates, ranging from 10 to 60 frames per second (fps) under two different scenarios: (a) the resource configurations set by a Deep Reinforcement Learning (DRL) agent versus (b) configurations determined randomly. As expected, the progression of data points indicates an increase in processing time corresponding to the increase in the number of fps irrespective of the strategies. It is observed that the DRL agent consistently achieves lower processing times across all frame rates. At the lowest observed frame rate of 10 fps, the average processing time of a frame with DRL is recorded at approximately 419 milliseconds (ms), in contrast to approx. 600ms when the configuration is determined randomly for the same fps. This pattern persists throughout the experiment under different fps rates. Such superior performance by DRL agent can be attributed to its ability to learn and adapt, unlike the random approach, which does not consider the system's performance feedback. The DRL agent likely sets suitable configurations in a way that is sufficient for the incoming frames, avoiding both excess and deficiency of resources. The random configuration method does not have such capacity to learn from the previous experience. A similar pattern is also observed when the number of fps increases to 60, where DRL is able to process in 20% less time compared to a random resource configuration.

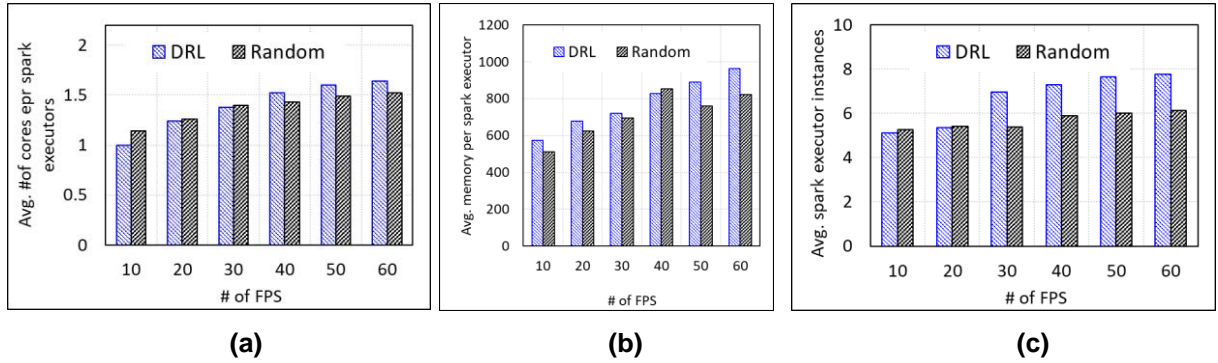

**Figure 7:** Average resource allocation with DRL method and without DRL or random approach (a) average number of cores assigned to each spark executor (b) average memory allocated to each executor and (c) average number of executor instances

Figure 7 provides insight into the overall resource configuration determined by DRL agent and random approach across all forms of experiments. The resource configuration mainly refers to the number of executor instances used, and the resources (CPU cores and memory) allocated to those executors. From Figure 7, we can deduce that as the number of fps increases from 10 through 60, there is a corresponding incremental increase in the number of Spark executors, cores and memory allocated by both the DRL agent and the random configuration.

Figure 7(a) shows the average number of Spark executor cores used for processing video frames under two different resource configuration strategies: Firstly, when the DRL agent sets it and secondly, when it is determined randomly. When the fps rate is set to its lowest number i.e. 10 fps, the DRL agent configures the system to use an average of one core per executor, while the random configuration sets the cores to an average of 1.14 cores per executor. As the frame rate increases to 60 fps, the DRL agent allocates an average of 1.65 CPU cores, in contrast to the random configuration, where an average of 1.52 cores are assigned to each executor. Similarly, the average memory in megabytes (MB) allocated to the executors under different fps rates is shown in Figure 7(b). In this figure a clear consistent increasing trend in memory allocation with the increase in fps rate is observed. Starting from approx. 600 MB for 10 FPS, the memory allocation rises consistently to approximately 1000 MB for 60 FPS as determined by the DRL agent. Such a pattern suggests that the DRL agent is trained to scale up memory resources in response to an increase in the number of frames being processed in each second. On the other hand, an increase in memory allocation with rising FPS is also observed when the configuration is decided randomly, but the increments do not appear as consistent or as proportionate as those of the DRL agent. For instance, at 10 fps, the random configuration begins at a lower memory allocation of approx. 500 MB and ends at approx. 800 MB for 60 FPS, which is significantly less than what is allocated by the DRL agent for the same fps rate. While the difference in the number of cores and memory allocation between the DRL and random configurations may appear small as in Figure 7(a) and 7(b), the impact on processing times can be significant, as shown in Figure 6. The time taken to process the streams of video frames is also affected by the number of spark executors created to process those frames. The average number of executor instances created randomly and by the DRL agent with varied fps rates is shown in Figure 7(c). It is observed that as the number of fps increases, the DRL-configured Spark executor instances also increase. Starting at approximately five executors for handling the incoming video at 10 fps, the number of executors rises steadily to more than an average of 7.5 when the fps rate increases to 60. This gradual increase suggests that the DRL agent actively scales up resources to accommodate the higher processing demands that come with increased frame rates, unlike the random approach. The random approach shows a less systematic and potentially less efficient approach where the required number of executors does not increase in direct proportion to the frame rate, resulting in suboptimal performance. The trends in Figure 7 suggest that the DRL agent is more sensitive to the increase in fps rate and thus consistently increases resources to handle the additional load, unlike the random approach.

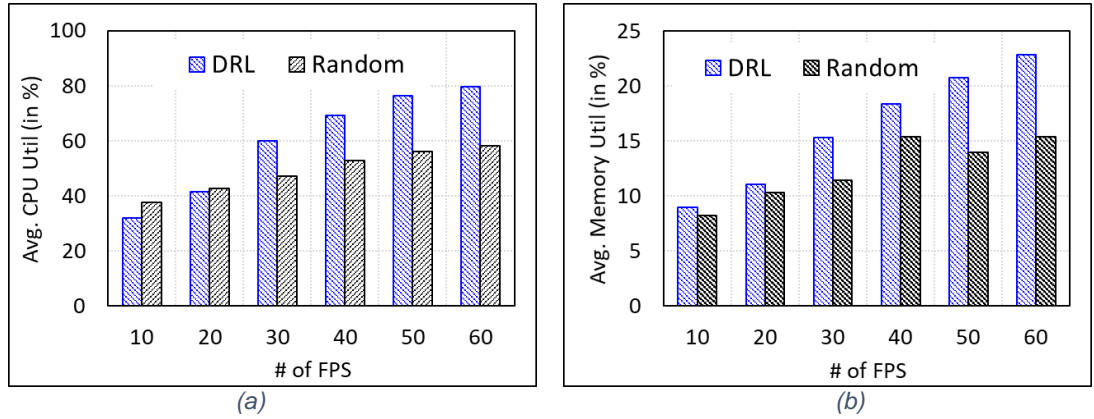

**Figure 8:** Average resource utilization with DRL method and without DRL or random approach (a) CPU cores utilization and (b) memory utilization

In addition to processing time and average resource configuration, as shown in Figures 6 and 7, we also observed the average resource (CPU cores and memory) utilization (in percentage) and presented in Figure 8. Figure 8(a) compares average cpu utilization between configurations set by a DRL agent and a random approach across varying numbers of fps. Similarly, Figure 8(b) shows the average memory utilization across varying numbers of fps rates. When the DRL agent is used to determine the resource allocation, the CPU utilization increases from approx. 30% to 80% and memory utilization increases from approx. 9% to 23%, as shown in Figure 8(a) and 8(b), respectively. On the other hand, under random configuration, a higher cpu utilization of approx. 37% and memory utilization of approx. 8% is observed when the fps rate is only 10. However, the same rate of increase in resource utilization is not maintained while the fps rate increases to 60, where CPU utilization is less than 60% and memory utilization is near to 15%. The results in Figures 6, 7 and 8 indicate the efficiency of DRL. It is able to make efficient usage of resources with the right set of configurations without compromising the processing time. Such comparison highlights the potential advantages of employing DRL for the management of resources in computationally intensive applications like video processing using Apache Spark.

### 5.4.2. DRL training performance

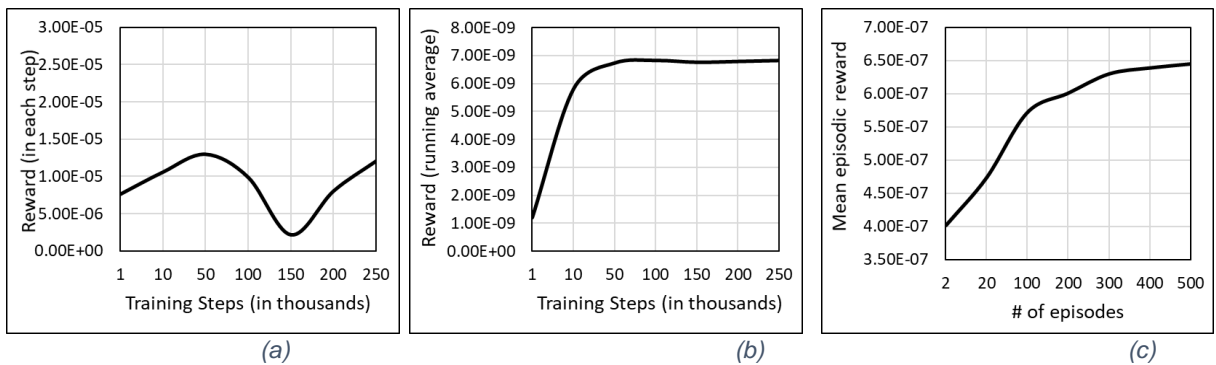

**Figure 9:** DRL training performance over time (a) reward received in each steps (b) Running average reward received by agent and (c) mean episodic reward received by the agent.

Figure 9 provides an overall performance of the DRL agent at different stages of its training process. The learning rate is also set to 0.001, which influences the rate at which the agent updates its knowledge base. The X-axis in all subfigures represents the number of steps (or iterations) the DRL agent has undergone in the training process. We carried out a total of 250 thousands steps with mix number of fps rate ranges between 10 and 60. With different steps and episodes, running reward average and episodic average of reward are calculated to observe the performance of the training process. A larger number of steps indicates the agent has explored and learned the environment better. The rewards in each step are inversely proportional to the processing time of the streaming frames. A higher processing time will lead to lower reward gain. Figure 9(a) shows the reward value the agent received in each step, which ranges between *2.10E-06* and *1.30E-05*. As shown in Figure 8(a), it is observed that agent received an initial lower reward value, which then increases, indicating that the DRL agent is starting to learn from its previous experience. As training continues, there is a noticeable fall in the reward, followed by a recovery. This is due to unexpected behavior of the cluster and other tools, including Spark, Kubernetes, log manager, and storage server during those steps. Such unexpected behavior of the entire setup could be due to the agent exploring less successful strategies or actions.

Similarly, Figure 9(b) shows the running average of the reward received by the DRL agent over training steps. Initially as training begins, the average reward is low, but it quickly rises as the DRL agent starts learning from its interactions with the cluster/environment. This initial rapid increase, until approximately 10 thousands steps, indicates that the agent is successfully identifying and exploiting strategies that yield higher rewards. The plateau phase, where the pattern flattens and average reward becomes stable or has no significant gains, suggests that the agent has reached a level of consistent performance and converged reward, which occurs approximately after 50 thousand steps. The plateau occurs because, as the agent learns more about the environment, the relative improvement in performance for each new discovery decreases. In other words, the agent has possibly identified and is now exploiting the most rewarding strategies available given the constraints and the dynamics of the environment. We also observed the mean episodic average reward received by the agent, as presented in Figure 9(c). The mean episodic average reward is also known as the average reward the agent receives for every episode, where an episode is a set of steps. This metric helps in understanding the agent's performance over longer sequences of actions, providing insight into how well the agent is learning to configure the resource for a specific fps rate. Each data point on the graph in Figure 9(c) corresponds to the average reward received by the agent per episode, with the number of episodes ranging from 2 to 500. Initially, the mean episodic rewards are lower, approximately 4.00E-07, and as the agent continues to learn and refine its strategies through thousands of interactions with the spark environment, these rewards gradually increase to approximately 6.5E-07 by the time the agent has completed all the training steps, e.i. 250,000 or 500 episodes. Similar to Figure 9(b), the mean episodic reward in Figure 9(c) also suggests the convergence of reward after approx. 250 episodes.

### 5.4.3. Computation overhead

Figure 10 illustrates the time taken by Apache Spark and the DRL agent during and after the training process. The results obtained demonstrate the benefits of using a DRL agent to optimize resource configuration at the cost of minimal additional computation time required by the agent. Figure 10(a) presents the computational overhead during the training phase of the DRL agent, comparing it with the overhead of the Apache Spark cluster. The x-axis represents the number of frames per second (fps), ranging from 10 to 60 fps, while the y-axis displays the computation overhead as a percentage. The observed results show that the DRL agent introduces a consistent computational load to the system during its training phase. It is noteworthy that the number of frames being processed per second by Spark does not influence this overhead. At 60 fps, the DRL agent accounts for less than 20% of the total training time, while Spark spends the remaining approximately 80% of the time processing the input frames. A similar pattern is observed at 10 fps, where the DRL agent's training consumes less than 15% of the time, and approximately 85% of the time is spent processing the input frames with Spark's resources configured by the DRL agent.

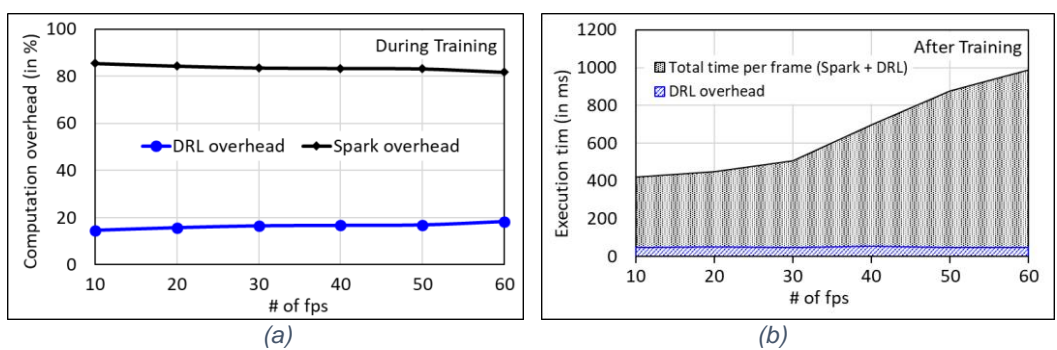

**Figure 10:** Time spent by DRL and Apache Spark (a) during training and (b) after training

In Figure 10(b), the execution time in the testing phase is observed. The graph shows a clear and expected trend where the total execution time, including both Spark and DRL processing, increases as the fps rate rises from 10 to 60 fps. For example, at 10 fps, the average time required to process a frame is approximately 400 milliseconds. This processing time climbs to nearly 1000 milliseconds when the system receives input at 60 fps. The results indicate that the DRL is responsible for adding approximately 45-55 milliseconds to the processing time, irrespective of the input fps rate. When the fps rate is set to 60, a 5% overhead is attributed to the DRL. Conversely, at 10 fps, a 10% overhead by the DRL is observed, demonstrating the DRL agent's consistent computational demand regardless of the frame rate.

## 6. Conclusions and Final Remarks

Heterogeneous data sources continuously generating data are becoming mainstream in today's ubiquitous world. Social media feeds, vital signs measurements, stock trades, and mobile call records are all examples of continuous data that is analysed by streaming applications to produce meaningful knowledge in real time. These insights are key for enterprises worldwide; hence, the ability to efficiently and automatically process this data as proposed by our vision will support these organizations and contribute highly towards the global economy. In particular, the outcomes of the realisation of our proposed vision include: (a) general serverless framework applicable to a range of SPEs deployed in cloud environments, (b) DRL agents capable of managing SPEs and their environments to achieve and maintain performance goals, (c) principles and solutions to the limited exploration ability of DRL agents in large scale distributed environments, and (d) a MADRL solution to coordinate the actions of agents to automate and improve resource management in distributed environments.

As billions of devices such as cars, farm machinery, and surveillance cameras will be connected to Internet by 2030 [20], this will create a digital era with extraordinary business opportunities for Australia. It is estimated that these devices will produce approximately 5 quintillion bytes of data everyday [21]. Approaches proposed in this article will enable the analysis of these data and the delivery of insights in a timely manner.

This vision article's objective is to propose the creation of Multi-Agent Deep Reinforcement Learning (MADRL) solutions for serverless stream processing systems with the goal of supporting the automated and efficient deployment of real time data analytics while ensuring Service Level Objectives (SLOs) are met and maintained throughout the applications' lifetime, thus driving commercial and scientific advancements. To achieve this objective, our future directions entail the following guidelines:

- Propose an architectural framework for Stream Processing Engines (SPEs) with the ability to automatically adapt their configuration to maintain SLOs in response to changes such as load spikes and performance variation.
- Define the actions, rewards, and states that would enable a multi-agent deep reinforcement learning system to efficiently manage the cluster resources while working towards achieving and maintaining a user-defined service level objective. Agents in the system will have the ability of interacting with the cloud environment and stream processing system by (a) scaling the cluster resources, (b) scaling the parallelism of topology components, (c) tuning system parameters, (d) dynamically configuring the placement of topology components in the cluster machines, and (e) dynamically adjusting the amount of resources assigned to containers.
- Investigate the problem of high exploration cost for RL agents in SPEs deployed in large-scale distributed platforms and the trade-off between limited or conditioned exploration vs. agent performance.
- Develop a software platform incorporating the above mechanisms and techniques for serverless management of resources in stream processing engines.